\documentclass[aps,prc,reprint,qshowpacs,groupedaddress,superscriptaddress]{revtex4-1} 

\usepackage[utf8]{inputenc}
\usepackage{multirow}
\usepackage{amsmath,amssymb,amsfonts}
\usepackage[pdftex]{graphicx}
\usepackage{hyperref}

\usepackage{subfigure}
\usepackage{units}
\usepackage{placeins}
\usepackage{xcolor}
\usepackage[super]{nth}
\graphicspath{{./images/}{./images/}}

\definecolor{blue}{rgb}{0,0,1}

\begin{document}


\title{Avoiding common pitfalls and misconceptions in extractions of
  the proton radius}

\author{Jan C.~Bernauer}\email{bernauer@mit.edu}\affiliation{Laboratory for Nuclear Science, MIT,  Cambridge, Massachusetts 02139, USA}
\author{Michael O.~Distler}\email{distler@uni-mainz.de}\affiliation{Institut f\"{u}r Kernphysik,  Johannes-Gutenberg-Universit\"{a}t Mainz, D-55128 Mainz, Germany}

\begin{abstract}
In a series of recent publications, different authors produce a wide
range of electron radii when reanalyzing  electron proton scattering
data. In the light of the proton radius puzzle, this is a most
unfortunate situation. However, we find flaws in most analyses that
result in radii around $\unit[0.84]{fm}$. In this paper, we explain our
reasoning and try to illustrate the most common pitfalls.
\end{abstract}
\maketitle




\section{Introduction} \label{Intro}
The term ``proton radius puzzle'' paraphrases
the disagreement between muonic hydrogen
Lamb shift experiments ($\unit[0.8409(4)]{fm}$)
\cite{Pohl:2010zza,Antognini:1900ns} and both atomic and scattering
experiments using electrons, summarized in the CODATA value of
$\unit[0.8751(61)]{fm}$ \cite{Mohr:2015ccw}.  The extraction of the
proton radius from scattering data is a treacherous business. In the
discussion about the proton radius puzzle, many pitfalls we and others
succumbed to became obvious. This paper is meant as an illustrated
guide of these.

The paper is divided in two main sections: in the first section, we
discuss missteps and misconceptions in general terms. The second
section discusses the flaws in the analysis of some recent papers.

\section{Comments on common mistakes and misconceptions}

In the following sections we discuss common mistakes that are
somewhat specific for the extraction of the proton radius from cross
section data (\ref{sec:taylor} - \ref{sec:rescale}). Starting with
section \ref{sec:tests} we talk about general properties of estimators
which are relevant whenever a given quantity is calculated based on
observed data.

\subsection{A polynomial fit is \textit{not} a Taylor expansion
  around 0, and the convergence is \textit {not} limited by cuts in
the time-like region.}\label{sec:taylor}
A polynomial in normal form, i.e., of the form 
$$ \mathrm{poly}(x,\vec{p})=p_0+p_1\cdot x+ p_2\cdot x^2+...$$ looks
identical to a Taylor expansion around 0:
$$ \mathrm{taylor}[f](x)=f(0)+\frac{1}{1!}\left.\frac{df}{dx}\right|_0\cdot x +\frac{1}{2!}\left.\frac{d^2f}{dx^2}\right|_0\cdot x^2+...$$
However, a fit of the polynomial does not yield the Taylor expansion.
This can trivially be seen just looking at the definition: a Taylor
expansion of a function around a point is given by the derivatives of
that function at that point. This necessitates that the function
indeed has these derivatives, and the value of the function at any
other point is of no consequence for the expansion. 
The polynomial used in a fit might look like a Taylor expansion, but
it is not: the coefficients of the polynomial are influenced by all
data points, i.e., it depends on the functional value at many ordinate
points.  
A fit with a polynomial written like a Taylor expansion around a different point $x_0$,
i.e.,\
$$ \mathrm{poly}(x,\vec{p})=p_0+p_1\times(x-x_0)+....$$
will find a different parameter vector $\vec{p}$, but transforming the
polynomial to normal form by multiplying out the parenthesis will
yield the same polynomial, independent of the choice of $x_0$.
It is worthwhile to note that the polynomial fit in general does not yield a
Taylor expansion at all, i.e., there is no common point $x_0$ where
the polynomial and the true function have the same value and
derivatives.

Indeed, according to the Weierstrass theorem, any function continuous
in an interval can be approximated to arbitrary precision and with
global convergence (over the interval) by a polynomial. This alone
does not guarantee that the first derivative is also approximated
well, the requirement for an accurate extraction of the
radius. However, it is trivial to show that this is true if the
function is continuously differentiable.

In contrast to the theorem of  Weierstrass, which concerns itself with
convergence of the maximum error, i.e., norm $\|\dots\|_\infty$, the
typical fit in the least squares sense minimizes according to norm
$\|\dots\|_2$, a fit-technical necessity (the error function needs to be
continous close to the optimal point) which also lends it itself to
the treatment of data with errors.

\begin{figure}
\includegraphics{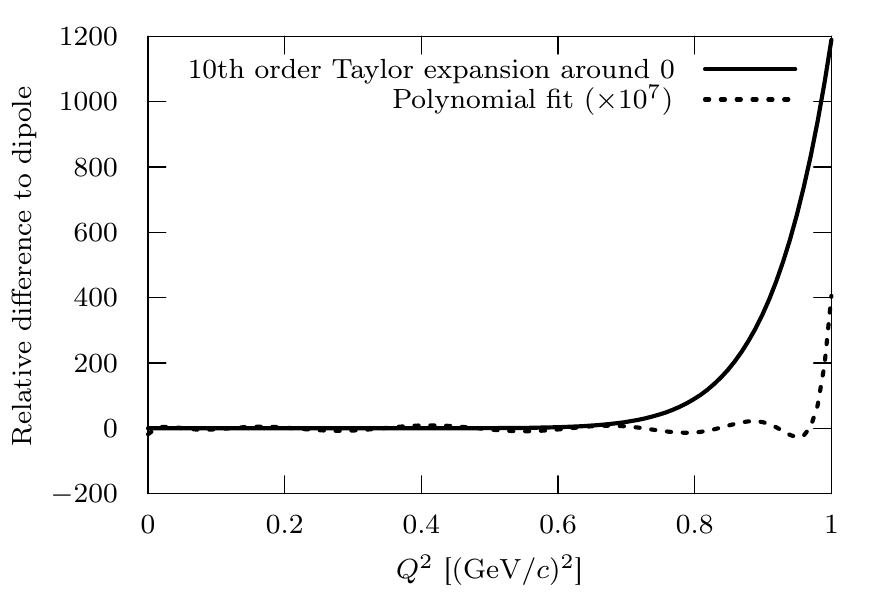}
\caption{\label{dipdiff}Relative difference of the polynomial fit and
  of a Taylor expansion to the dipole, as a function of $Q^2$. Please
  note that the difference of the polynomial is scaled up by a factor
  of 10 million, i.e.\ the difference is less than 40 ppm for the whole
  displayed range.}
\end{figure}

The prevalence of the notion that a polynomial fit is somehow related
to a Taylor expansion is striking
\cite{Higinbotham:2015rja,Griffioen:2015hta,Horbatsch:2015qda,Lorenz:2014vha,Paz:2011qr,Hill:2010yb}.
We want to present here an example: to this end, we generated $G_E$
values following the standard dipole, i.e., a dipole with a parameter
of $\unit[0.71]{(GeV/c)^2}$, at the $Q^2$ points of the Mainz data
set. These values are then fit with a \nth{10} order polynomial. The data
points are error-free, but we weight the points according to the
uncertainty present in the Mainz data set.  In Fig.\ \ref{dipdiff},
the difference of the polynomial fit and of a Taylor expansions around
$\unit[0]{(GeV/c)^2}$ truncated to \nth{10} order.  The standard dipole
has a pole at $Q^2=\unit[-0.71]{(GeV/c)^2}$, therefore a Taylor
expansion around 0 is limited in its convergence to a radius of
$\unit[0.71]{(GeV/c)^2}$. As expected, the Taylor expansion diverges
strongly from the dipole close to $\unit[0.71]{(GeV/c)^2}$. In
contrast, the polynomial fit does not diverge form the dipole by more
than 40 ppm between $0$ and $\unit[1]{(GeV/c)^2}$. Indeed, the
polynomial fit approximates the dipole better than the Taylor
expansion for all $Q^2$ above $\unit[0.15]{(GeV/c)^2}$.

Many authors \cite{Higinbotham:2015rja,Griffioen:2015hta,Horbatsch:2015qda,Lorenz:2014vha} argue that a polynomial fit is limited in its
convergence to $Q^2<4m_\pi ^2$ because of a pole
at $Q^2=-4m_\pi^2$, i.e., in the time-like region, and limit their
fits to the region below, even for non-polynomial fits. As shown, this
reasoning is wrong.

\begin{figure}
\includegraphics{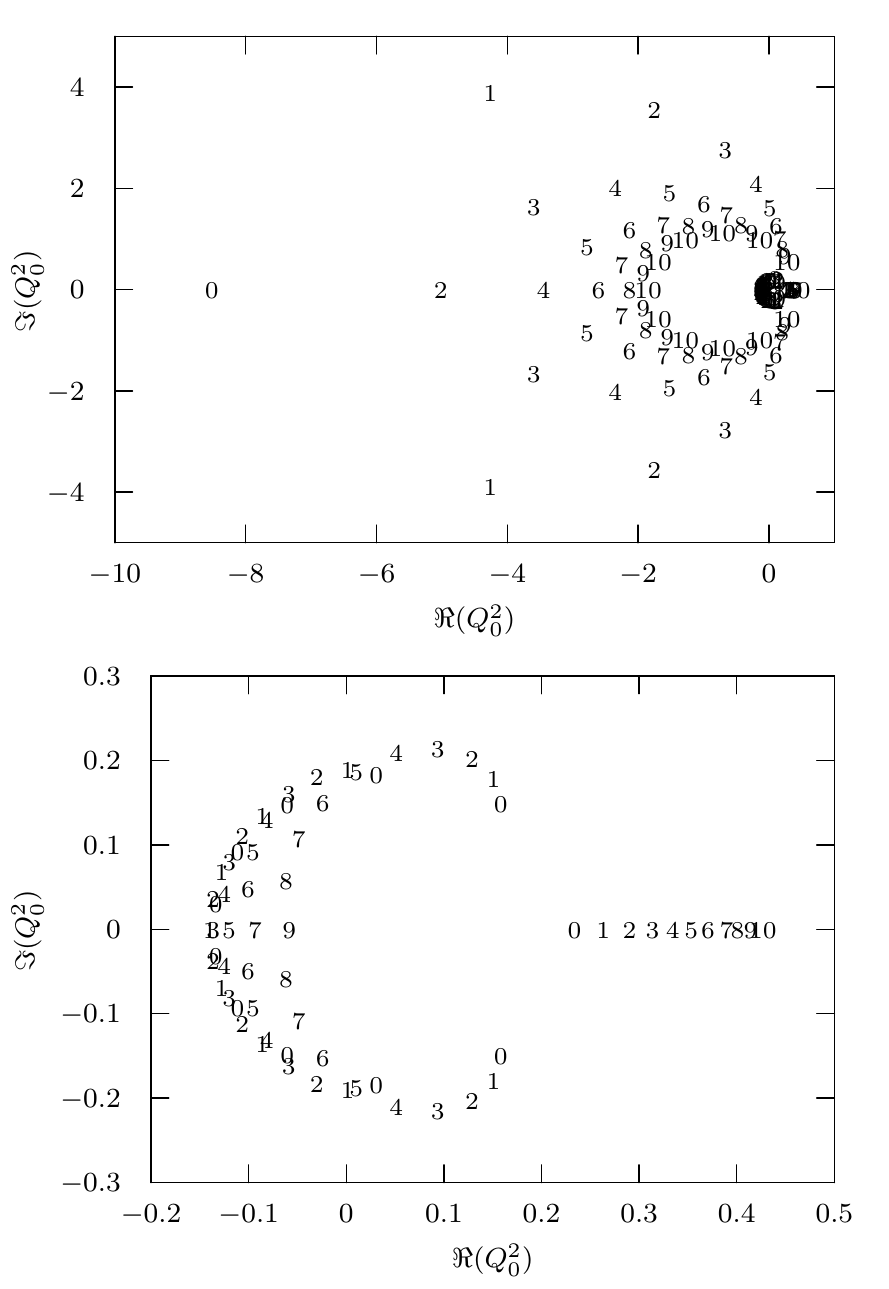}
\caption{\label{dipzero} A comparison of the polynomial fit
  coefficients with the symbolic expression for a Taylor expansion of
  the dipole at an arbitrary $Q^2_0$ order-by-order yield 12
  complex-valued $Q^2_0$ for each order, displayed here in the complex
  plane and labeled by the order they stem from. None of them coincide for all
  orders, proving that the polynomial fit is not a Taylor expansion of
  the dipole at all.}
\end{figure}

Of course, a Taylor expansion around a $Q^2_0$ more centered in the
$Q^2$ interval one is interested in would perform better.  One might
be led to believe that the fit might relate to a Taylor expansion not
around 0, but around a $Q^2_0\neq 0$, an effective, weight-averaged
center of gravity of the data points (indeed, the authors held this
believe briefly). But this is not true in general, as can be shown for
this example. From the coefficients found in the fit, one can
calculate, order by order, which possible $Q^2_0$ these belong to. In
the example case at hand, one each order, one finds 12 possible
$Q^2_0$, however, none of them are common to all orders, as is
illustrated in Fig.\ \ref{dipzero}. Therefore, the best fit polynomial
is not a truncated Taylor expansion of the dipole function around any
(one) point.

\subsection{Unconstrained fits with conformal mapping
is \textit{not} a good idea}\label{sec:conformal}

Conformal mapping is used by some to avoid the perceived problem of
the convergence radius.  E.g., in \cite{Lorenz:2014vha}, the authors define the function 
\begin{equation}
z(t,t_\mathrm{cut})=\frac{\sqrt{t_\mathrm{cut}-t}-\sqrt{t_\mathrm{cut}}}{\sqrt{t_\mathrm{cut}-t}+\sqrt{t_\mathrm{cut}}},
\end{equation}
with $t=-Q^2$ and $t_\mathrm{cut}=4M_\pi^2$. The form factors are then expressed as a polynomial in $z$ instead of
$Q^2$,
\begin{equation}
G_{E/M}(t)=\sum_{k=0}^{k_\mathrm{max}} a_k\cdot z(k)^k
\end{equation}

 The mapping function maps the whole positive $Q^2$ range into
the range $[0..1]$ in a rather non-linear fashion, compressing the
larger $Q^2$ values to a very small range in $z$ close to 1. On the
other hand, the very-low $Q^2$ is mapped to a comparable large
range. To illustrate this point further, in the unmapped case, the fit
has to ``bridge''  from $0$ to $\approx\unit[0.004]{(GeV/c)^2}$, or
about $0.4\%$ of the range of the data. In the mapped case, it has to
bridge from $0$ to $0.0133$, or about $2.2\%$ of the range of the data.
It follows that the flexibility of the polynomial expansion is
shifted to the low-Q range, which leads to multiple problems:
\begin{itemize}

\item In the low-Q region, the fits are very flexible. However, the
  data starts at a minimal $Q^2$, so that a fit can introduce arbitrary
  structures below the data. The extraction of the radius from the
  data is only meaningful if one assumes that such structures do not
  exist. This is warranted, as such structures typically lead to a
  charge density distributions with pathologically large densities at
  large radii \cite{Sick:2012zz}. 
  Additionally, and even more relevant here, is that the analysis
  extracts both electric and magnetic form factor at the same
  time. The large flexibility of the model makes this completely
  unstable, as we show below. This also influences the charge radius
  extraction. At the lowest $Q^2$, there are only measurements for
  one beam energy, and a Rosenbluth separation is not possible. A
  extraordinary flexible model for $G_M$ in that region can ``steal''
  from the electric form factor, affecting the extracted radius.

\item The compression of the larger $Q^2$ to a small range of $z$ values
  exacerbates a problem inherited by many polynomial-type fits: the
  parameters tend to get very large, but the contributions to the fit
  of the different orders cancel to a large extend, especially at
  large z. At small z, only a small difference remains which is
  exploited by the fit algorithm to explain the data. However, many
  combinations of large parameter values exist which all give similar
  quality of fits, but are far apart in parameter space. Care must be
  taken that the fit actually converges to the best minimum. 

\end{itemize}
Both of these points can be somewhat addressed by constraining the
parameters, as has been carried out in \cite{Hill:2010yb} for an older data
set. On the other hand, in a fit to the Mainz data, Lee et al.\
\cite{Lee} find a strong dependence on the cut-off in $Q^2$. We
believe this to be a consequence of aforementioned points.

\subsection{A good $\chi^2$ does \textit{not} signal
  a trustworthy extraction of the radius}\label{sec:chisqr}
To rely on $\chi^2$ to indicate a good fit is dangerous. In the
original meaning, it is a test of the data quality; assuming that a)
the model is correct, b) the errors are statistical and exactly known and c) the individual
data points are independent (or their correlation is at least known), it expresses how likely it is that the data are drawn from the distribution given
by the model. 
All of these assumptions are typically violated:
\begin{itemize}
\item One normally does not know whether it is the correct
  model. Indeed, this is what one wants to test.  An incorrect model,
  however, can produce small $\chi^2$ values and still be wrong.
\item In many experiments, especially the Mainz data set, a
  sufficiently large part of the errors is not driven by counting
  statistics but other effects. This limits the knowledge we have
  about the errors.
\item Data have systematic errors which couple the data points. The
  summands in the $\chi^2$ sum are not independent, but the correlation is unknown.
\end{itemize}
We refer to Kraus et al.\ \cite{Kraus}, for an illustrative discussion.

One more caveat: the minimal sum of the weighted squares
of deviations of the data from a model function should be
distinguished from $\chi^2$ and we usually call it $M^2$. For the
reasons given above, $M^2$ does not follow a $\chi^2$-distribution in
general. However, we will adhere to the common practice and call it
$\chi^2$ in the following chapters.

\subsection{Low-order fits are \textit{not} a good idea}\label{sec:loworder}
While one would hope that a linear model converges to the same value
as a higher order model if the $Q^2_\mathsf{max}$ is suffiently small,
the current state of the data clearly does not reach far enough
down. We again refer to Kraus et al.\ \cite{Kraus} which discuss this
at length.
As an additional caveat, we want to repeat that the polynomial fit is
not a Taylor expansion. In a truncated Taylor expansion, the error at the
expansion point is zero, and grows from there. One expects that a
lower-order expansion has a smaller radius in which the error is below
a certain threshold, but the error is still zero at the expansion
point. However, in a fit, this is not true. While a lower-order fit
will have a bigger error, the localisation of the error is less
clear. A fit will approximate the local slope of the data (i.e., at
$Q^2>0$), not at 0.

\subsection{Common fit algorithms do \textit{not} always
find the true minimum}\label{sec:minimum}
In a fit, one searches for the global minimum of $\chi^2$, the
absolute best parameters. Depending on the particular model, the
$\chi^2$ landscape can have many local minima, and many fit algorithms
are prone to get stuck in one of them. In our fits, we found that both continued
fraction expansion and conformal mapped polynomial type fits are
especially susceptible to this problem.
Except for an exhaustive search, which is prohibitively slow, there are no generally robust 
algorithms available, but simulated annealing is often succesful even
in hard cases. 
In our fits, we test for this problem by fitting repeatedly with
different, random start values. This can help find a better minimum in
many cases, however it's impossible to prove that the found minimum is
indeed the global one. We recommend to avoid models which have too
many local minima; depending on the noise in the data, the true
minimum might not be the global minimum using that particular data
set.

Another indication for this type of problem is the dependency of
$\chi^2$ on the fit order $N$.
 For any group of models $G_N$, where the images in function space,
\begin{equation}
\mathfrak{I}(G_N)=\left\{G_N(Q^2,a_0,a_1\ldots
  a_N),\ \forall a_k\in \mathbb{R}\right\},
\end{equation}
fulfill the relation
\begin{equation}
\mathfrak{I}(G_N)\subseteq \mathfrak{I}(G_{N+1}),
\end{equation}
the $\chi^2$ achieved by the models must monotonically decrease as a function of
$N$:
\begin{equation}
\label{chidecrease}
\chi^2_{N+1}\leq\chi^2_N
\end{equation}
Before we implemented the randomized start value approach from above,
fits of polynomial models violated this condition when the number of
parameters was excessively large.

\subsection{Rescaling the errors in the Mainz data set does
  \textit{not} allow for bad fits to be correct}\label{sec:rescale}
In the Mainz analysis \cite{Bernauer:2010wm,Bernauer:2013tpr}, we use
the $\chi^2$ of our best model to determine the size of point-to-point
errors on top of the counting statistics errors. This might
overestimate the errors in two ways; the data also contains systematic
errors, and even the best model might have systematic differences from
the true model. On the other hand, the model might overfit the data,
giving a slight underestimate of the errors. In total, we believe the
errors to be accurate to $<10\%$.  Many take this as a license to scale the errors
up if their fit produces a too large $\chi^2$. Doing so, however,
would not change the relative ordering of the fits; the better fitting
models still are better, and an explanation for the worse fit of their
model must be given.

\subsection{A statistics test can \textit{not} tell
  which model is the true model}\label{sec:tests}
When fitting data where the true model shape is unknown, as is the
case for form factors, we must resort to flexible models like
polynomials or splines. The crucial question is now how flexible the
model actually has to be---one has to balance between minimizing bias
and possible overfitting.  One is tempted to try to deduce from the
data how much flexibility is needed, and indeed we do the
same. However, one has to be very careful: typical statistical tests,
like the F-test, are used to identify a model that best fits the
data. It can not prove that the simpler or the more complicated model
is true, nor that the parameters it extracts are unbiased. For an
example, see Section \ref{sec:occamsbeard}.

Additionally, one has generally an interpretation problem: in the
standard F-test, the zero hypothesis H0, which one tries to disprove, is: the
simpler model is correct. The rest of the method now assumes H0 to be
correct, tests whether the data conforms to that and based on this
rejects or accepts H0. To this end, one defines a false rejection
threshold, i.e., one finds a threshold for the test function so that
one would falsely reject H0 even if it's true with a small
probability.
However, this is decidedly not related to the probability that H0 is
actually correct, because one does not know how often the test would
accept/reject H0 if H0 is actually false.

The falsehood of the approach can be illustrated differently: taking a
large data set, one finds that a given complexity is advocated by these
methods.  Reducing the data set, for example by a $Q^2$
cut-off, will require a simpler model. However, in truth, only one (or
none) of these models can be true, invalidating the theoretical basis
of the test.  

For nested problems, in general, the coefficient of a lower order
changes when higher orders are fitted. While the data might not be
good enough to prove that these higher orders are required, they might
still be there, and neglecting them in the fit leads to a bias. For
polynomials, one can find a basis orthogonal in respect to the data,
for example via the Forsythe method. Then, indeed, one can use a
statistical criteria to select the number of basis functions without
affecting the extraction of quantities related to the lower order
coefficients. Unfortunately, the radius, i.e., the linear term, appears
in all orders except for the constant term, so that this approach does
not help for the problem at hand.

For purely polynomial fits, however, it is easy to see that any
hypothesis which truncates the order must be wrong: a polynomial will
always go to $\pm\infty$ for $Q^2\rightarrow\infty$, but we know that
the form factors approach 0. This means that any statistical approach
which assumes any truncated hypothesis to be true is built on sand. As
a consequence, the radius extracted with a truncated polynomial will
always have a bias from that truncation. This does not mean all hope
is lost, as this error gets smaller if one includes higher orders, a
consequence of the theorem of Weierstrass.

\subsection{An estimator is \textit{not} guarantied
  to be consistent and unbiased}\label{sec:bias}
It is necessary to review what it means in statistical terms to
indirectly ``measure'' a quantity like the charge radius given a set
of data, e.g., cross section data. We will stick to the frequentist
interpretation of statistics laid out in
\cite{James:2006zz,Agashe:2014kda} where probability is interpreted as
the frequency of the outcome of a repeatable experiment. None of the
following insights are new or original but can be found in many text
books on statistics. Most of the time we are only paraphrasing.

An indirect measurement translates to an estimate of a parameter.  An
estimator $\hat{a}$ is a function of the data used to estimate the
value of the parameter $a$. Therefore the estimator $\hat{a}$ is
treated like a random variable. As there is no general rule on how to
construct the estimator, one chooses a function with optimal
properties. Important properties are consistency and unbiasedness
which relate the estimator $\hat{a}$ and the {\em true} value of the
parameter, $a_0$.

An estimator is called consistent if the estimator $\hat{a}$ is equal
to $a_0$ in the limit of an infinite
sample size:
$$\lim_{n\rightarrow\infty} \hat{a} = a_0$$
The bias $b$ of an estimator is the difference between the expected
value of $\hat{a}$ and the true value of the parameter:
$$b= E[\hat{a}] - a_0$$

Commonly used methods to construct such an estimator are the least
squares method or the more general maximum likelihood method. However,
there are many more possible methods to construct an estimator. Also, it
can not be implied that the method of least squares results in a
consistent and bias-free estimator, not even in the simplest
cases.

For example, given $N$ data points $x_i$, where $i=1,2,\ldots,N$ and
we assume the data points are drawn from a Gaussian
distribution. Using the maximum likelihood method one gets the
estimators for the mean and the variance:
\begin{eqnarray}
\bar{x} &=& \frac{1}{N} \sum^N_{i=1}x_i \label{eq:biasedmean} \\
s^2 &=& \frac{1}{N} \sum^N_{i=1}(x_i-\bar{x})^2 \label{eq:biasedvariance}
\end{eqnarray}
However, the maximum likelihood estimator for the sample variance in
equation~(\ref{eq:biasedvariance}) is biased. In this special case a
small change leads to the well known, bias-free estimator of the true
sample variance:
\begin{equation}
s^2 = \frac{1}{N-1} \sum^N_{i=1}(x_i-\bar{x})^2 \label{eq:unbiasedvariance}
\end{equation}

 Under very controlled circumstances, linear models, knowledge
of the true model function, known statistical errors, one can rely on
asymptotic properties. For all other cases a simulation with pseudo
data has to be performed in order to check the consistency and the
unbiasedness of the estimator used.

\subsection{The robustness of an estimator is \textit{not} self-evident}
\label{sec:robust}
The robustness of an estimator describes the insensitivity of the
estimator in the face of false data and false assumptions. In the case
of the proton radius extraction we want that our estimate is not
unduly affected by systematic errors in the data or the specific
functional form of the form factors that we use. Also the precise
value of the $Q^2$ cut-off or small changes in the cut parameter
(conformal mapping) should not affect the estimation.

To illustrate the importance of the robustness criterion we can
examine the estimation of the centre of an unknown, symmetric
distribution. As shown in many textbooks (e.g. \cite{James:2006zz})
the well known sample mean is only optimal if the distribution is
normal. For the double exponential distribution the optimal estimator
is the median and if the distribution is unknown one should use the
trimmed mean where the highest and lowest values of the sample are
removed and the sample mean is calculated from the remaining 46\% of
the observations.

This demonstrates that the robustness of an estimator is not at all
self evident, not even in the simplest cases. The properties of an
estimator have to be studied carefully.

\subsection{An estimator is \textit{not} necessarily efficient}
\label{sec:efficiency}
Recall that the estimate $\hat{a}$ of a parameter $a$ itself is a
random variable. We have discussed the bias of an estimator in
Section~\ref{sec:bias}. Now, we focus on the efficiency. In
statistics, the efficiency is about the variance of an estimator. An
efficient estimator has the optimal (minimal) variance. Again there
are very simple textbook examples where the standard procedure does
not provide the most efficient estimator.

Consider the mean of a sample: if the underlying distribution is the
uniform distribution, the use of eq.~\ref{eq:biasedmean} will not give
you the most efficient estimate of the sample mean. However, the
midrange
\begin{equation}
\bar{x}=\frac{\hat{x}+\check{x}}{2} \label{eq:midrange}
\end{equation}
which is the mean of the two extreme values within the sample has the
minimal variance. The arithmetic mean, which is the most efficient
estimate of the sample mean for the normal distribution, does poorly
for the uniform distribution. The variance of the estimator scales
with $1/n$ where $n$ is the sample size. When using {\it midrange} on
a sample drawn from a uniform distribution the variance is
proportional to $1/n^2$.

Again, a simulation with pseudo-data will help to evaluate the
variance of the estimator that is used.

\section{Comments on recent papers}

\subsection{Failed fits}
In the recent paper \cite{Lorenz:2014vha}, the authors use the
conformal mapping approach to fit the recent high precision form
factor data from Mainz \cite{Bernauer:2010wm,Bernauer:2013tpr},
claiming a 3 sigma reduction in the proton radius puzzle. We believe
that this finding is in error on multiple accounts: the fit function
is, as is, not suited to analyze the data, their fitting program does
not converge to the minimal solution, and their statistical approach
is flawed. Additionally, the comparison with the Mainz fits is not on
equal footing.

We tried to replicate the approach followed by Lorenz et al.\ in
\cite{Lorenz:2014vha}. Our results however differ significantly from
the ones reported there. The nature of the differences mainly point to
a failure of the fitting algorithm used in \cite{Lorenz:2014vha} to
reliably find the true minimum.  Trying to reproduce Fig.\ 1 of
\cite{Lorenz:2014vha}, 
we find a completely different $\chi^2$ evolution, namely significantly lower
values for smaller $k_\mathrm{max}$, even with a naive implementation of the
fitting routine.  

The original paper is not clear on whether $a_0$ is set to 1 or
fitted. Since fitting it would constitute a renormalization, we set
$a_0=1$. In any case, this limits the flexibility of our model, that
is, a fit including $a_0$ as a free parameter will produce an even
smaller $\chi^2$. 

For larger $k_\mathrm{max}$, we have to employ the more advanced
fitting algorithm described in Section \ref{sec:minimum} and find consistently lower numbers than what was reported in
\cite{Lorenz:2014vha}. 
Since they use a polynomial fit, the $\chi^2$ have to follow eq.\
\ref{chidecrease}, which is violated for $k_\mathrm{max}=13$ or $14$
(worse fit than $k_\mathrm{max}=12$).

\begin{figure}
\includegraphics{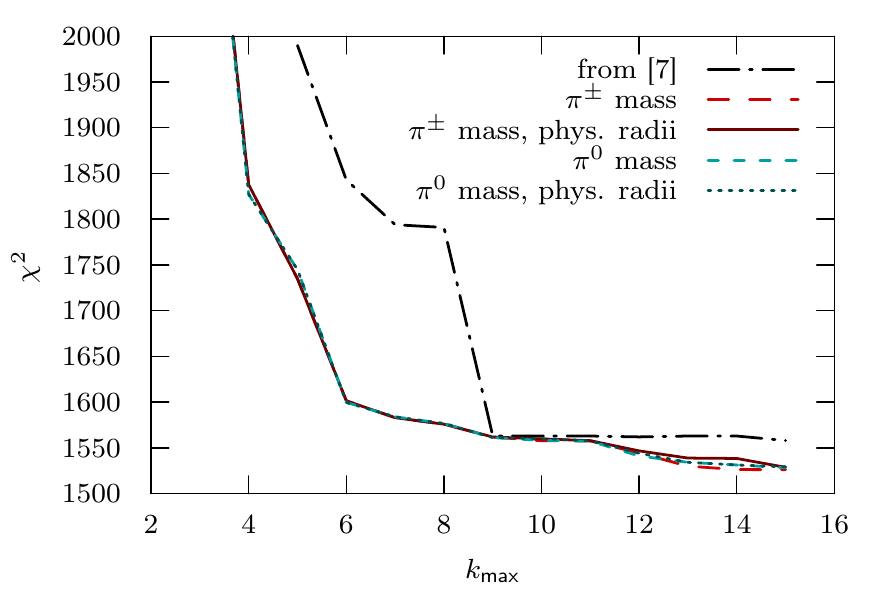}
\caption{\label{fig1}Comparison of the achieved $\chi^2$ in
  \cite{Lorenz:2014vha} and by us as a function of
  $k_\mathrm{max}$. Using the same data and fit function, we find
  substantially smaller $\chi^2$, with the characteristic knee at 6
  instead of 9.}
\end{figure}

We also find a rather strong dependence on the pion mass used in the
mapping and we therefore report both results (see
Section~\ref{sec:robust} on the robustness of an estimator). Figure
\ref{fig1} shows a comparison of our results and the ones from
\cite{Lorenz:2014vha}.  In Section~\ref{sec:robust}, we emphasized the
importance of the robustness of a model that is used to extract a
parameter like the charge radius from a set of data. The strong
dependence on the pion mass clearly violates that criterion of a good
estimator.
 
\begin{figure}
\includegraphics{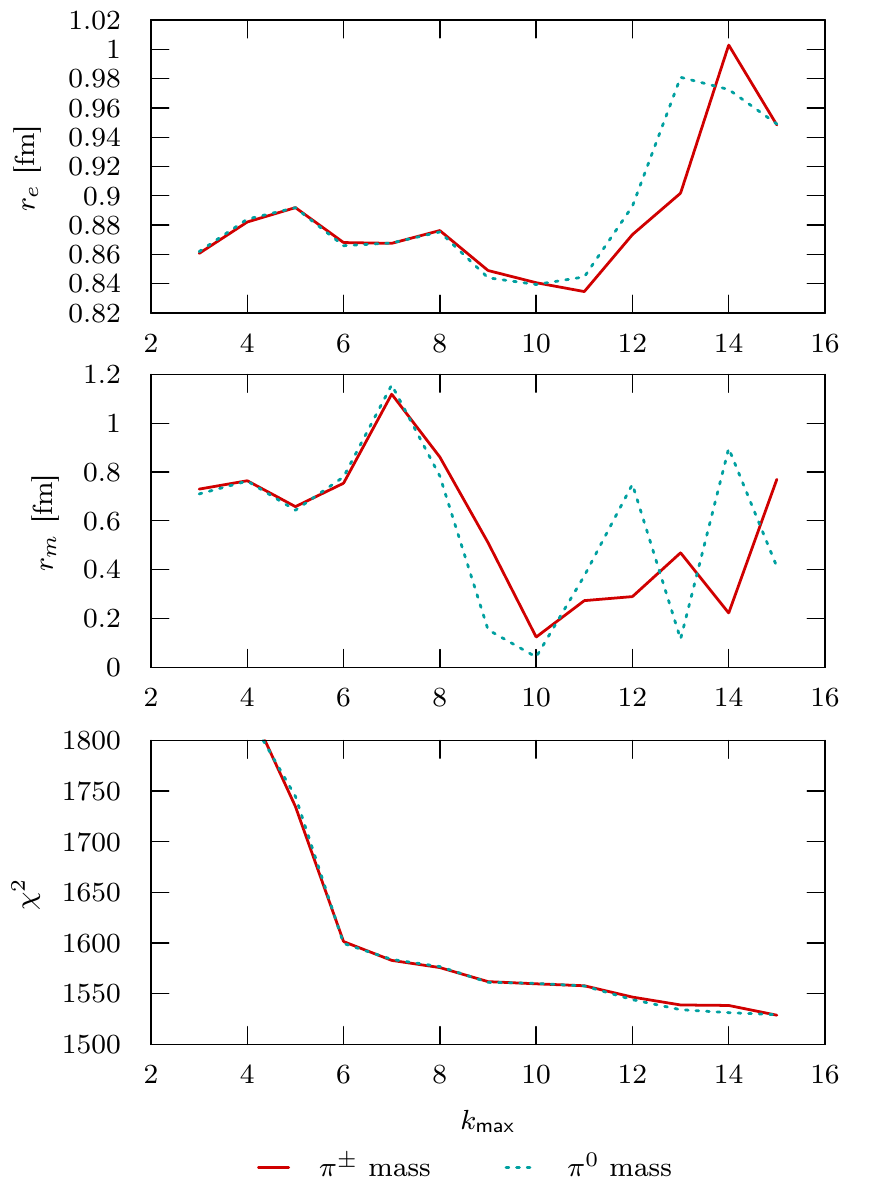}
\caption{\label{fig2}Extracted radii and achieved $\chi^2$ as a
function of $k_\mathrm{max}$, using a polynomial fit and conformal
mapping. The magnetic radius does not show a stable region for
$k_\mathrm{max}$ above the knee and we would therefore reject the
model all together.}
\end{figure}

For $k_\mathrm{max}\geq10$, the best solution found by the fit
sometimes produces a non-physical, that is, imaginary, magnetic radius. We
therefore also keep the best solution with a real magnetic radius,
which has a slightly larger $\chi^2$, still below the values found in
\cite{Lorenz:2014vha}. Both curves are shown in Fig.\ \ref{fig1}. We
only show results for valid radii in Fig.\ \ref{fig2}, which shows the
dependence of the extracted radii and reached $\chi^2$ on $k_\mathrm{max}$.

We find the typical ``knee'' in $\chi^2$ around $k_\mathrm{max}=6$, much
smaller than $k_\mathrm{max}=9$, found in
\cite{Lorenz:2014vha}. Compared to the fits in
\cite{Bernauer:2010wm, Bernauer:2013tpr}, the knee is softer, with visible reduction in
$\chi^2$ beyond the knee. We interpret this as a sign
that the fit is already overfitting the low-$Q^2$ region, but still
can make use of the added flexibility at larger $Q^2$, where the
mapping function compresses the range.

In contrast to \cite{Lorenz:2014vha}, we do not observe any stable plateau of the
radii. From the properties of the fit function, this is somewhat
expected. We can only speculate over the exact nature of what
caused the plateau in \cite{Lorenz:2014vha}. 

Following the procedure of the Mainz analysis \cite{Bernauer:2010wm, Bernauer:2013tpr},
we make use of two criteria to find suitable parameter numbers. The
lower bound is given by the position of the knee, while the upper
bound is found by looking for a plateau in both charge and magnetic
radii. The rational behind this is easy to understand: the knee
signals that the model has enough flexibility to follow the underlying
shape of the data. With less flexibility, the fit has a common-mode offset from
the data, leading to a large increase in $\chi^2$. With more
flexibility, the fit starts to follow local, statistical fluctuations,
which only reduce $\chi^2$ slightly. With further flexibility, the fit
gets unstable, which can be seen in the radii. Of course, these rules
are not rigorous, but constitute a good guide line for the selection.

As shown in Fig.\ \ref{fig2}, there is clearly no plateau in the
magnetic radius. We would therefore not accept the model at all.

However, it is interesting to note that, ignoring the magnetic radius for a
moment and focusing on the charge radius, the fit extracts values
in the range from $0.866$ to $\unit[0.876]{fm}$ for $k_\mathrm{max}=6...8$,
slightly lower, but in good agreement with our reported results.

\subsection{Low order polynomial fits to low-Q data}
\label{pseudodata}

Motivated by the perceived connection of polynomial fits to Taylor expansions
and their radius of convergence (see Sections \ref{sec:taylor} and \ref{sec:loworder}),
Griffioen et al.\ \cite{Griffioen:2015hta} fit first
and second order polynomials to the data up to
$Q^2=\unit[0.2]{(GeV/c)^2}$ and report radii close to $\unit[0.84]{fm}$.

To illustrate the problems of these fits, we generate two groups of
pseudo-data.  The first groups are generated from the \nth{10} order polynomial
fit from \cite{Bernauer:2010wm, Bernauer:2013tpr}, corresponding to a
radius of $\unit[0.8855]{fm}$, the other from a \nth{10} order polynomial fit to
the data of \cite{Bernauer:2010wm, Bernauer:2013tpr}, with the radius
forced to $\unit[0.841]{fm}$. For each group, we simulate 2000
repetitions of the Mainz experiment, generating 2000 data sets. 
These pseudo-data set, and the real data set, can now be analyzed in various
ways and one can compare the behavior of the fits to the real data
and to the pseudo-data sets. 
For the real data set, we selected the normalization using the
polynomial fit (see explanation in \cite{Bernauer:2013tpr}), and
use the (fixed) \nth{10} order polynomial fit for $G_M$ together with the
to-be-optimized model for $G_E$ to fit on the cross section level.

\begin{figure}
\includegraphics{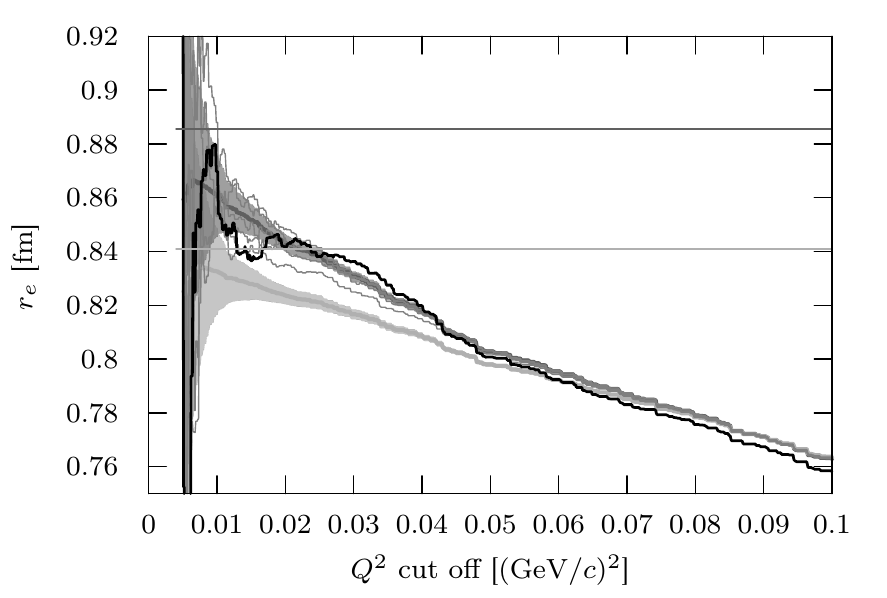}
\caption{\label{fofit}Extracted radii from linear fits to pseudo and
  real data, as a function of the $Q^2$ cut-off.  Black curve: fits to
  data; grey thick curves: average extracted radius to pseudo data
  (darker, upper curve: pseudo-data with large radius; lighter, lower
  curve: pseudo-data with small radius); bands around these curves are
  one-sigma point-wise error bands; dark grey thin curves: fits to the
  first five pseudo-data sets with large radius.
} 
\end{figure}

The results for the first order fits are shown in Fig.\
\ref{fofit}. The strong bias (see Section~\ref{sec:bias})
in the fits to pseudo-data is obvious,
even for very small cut-offs. At $\unit[0.02]{(GeV/c)^2}$, we find an
average bias of more than $\unit[0.04]{fm}$, yielding essentially the
small muonic radius of $\unit[0.84]{fm}$ despite having a true radius
of $\unit[0.8855]{fm}$.  It follows that results from linear fits are
unreliable; assuming that our polynomial fit is indeed an accurate
representation of reality, the bias observed for pseudo-data explains
the small radius found by Griffioen et al.\ \cite{Griffioen:2015hta}.

It is striking how similar the fit to data is compared to the fit to
the pseudo-data with large radius. It is worthwhile to note that the
real data may very well have systematic errors affecting
small groups of data points, not reflected in the generation of pseudo
data here.

\begin{figure}
\includegraphics{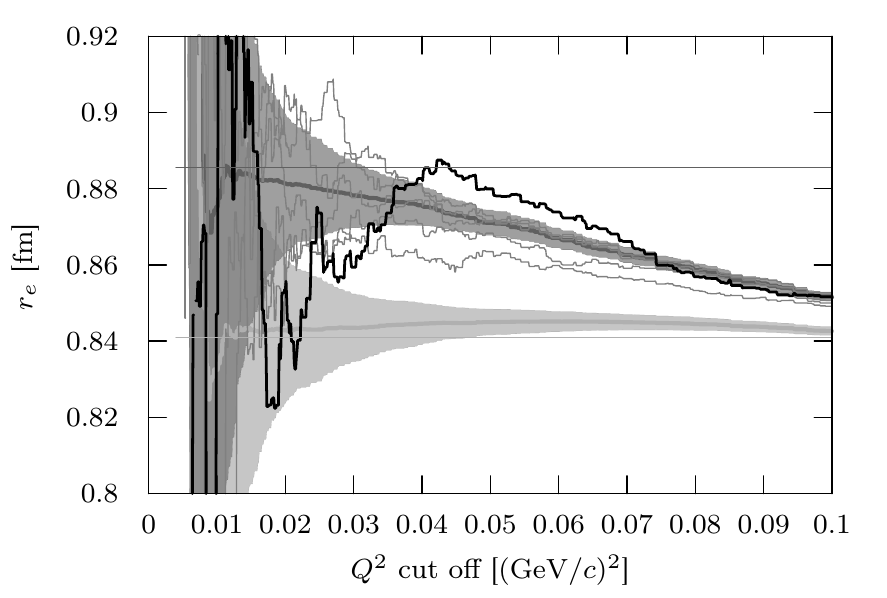}
\caption{\label{sofit}Extracted radii from quadratic fits to pseudo and
  real data, as a function of the $Q^2$ cut-off.  Curves as in Fig.\ \ref{fofit}.
} 

\end{figure}

A second order fit does somewhat better, as shown in Fig.\
\ref{sofit}. The overall picture is somewhat similar to the first
order fit. However, on average, the quadratic fit should have a much
smaller bias. Nevertheless, the errorband which is a measure of the variance is much bigger compared to the linear fits (see Section~\ref{sec:efficiency}).
The fits to data show a dip around the cut-off used by
Griffioen et al.\, but recover and come back to higher values, until
the bias lowers the extracted value again. Comparing the cut-off dependence of the fits to data to
that of fits to individual sets in the pseudo-data, one can see
similar swings, albeit maybe somewhat less pronounced. This might be
simply the result of a statistical fluctuations, or of a local problem
in the data around $\unit[0.02]{(GeV/c)^2}$. Both possibilities will hopefully
be addressed with future data. A low order fit to small data sets will
statistically be more sensitive to such perturbations: first, problems
at the highest accepted $Q^2$ will affect the highest order most, and
the effect is diminished on the first order term, more so if more data
are fitted with higher-order functions. Second, even assuming that the
probability that a data point is affected by such systematic effects
is constant (the probability is likely smaller for higher $Q^2$ data,
as corrections, e.g., due to backgrounds, are smaller), multiple 
systematic effects  in the larger data sets will partially cancel, so
their relative influence is likely proportional to $1/\sqrt{N}$.

For third order fits, Griffioen et al.\ propose to
expand the form factor as
$$G_E(Q^2) = 1 - \frac{1}{6} R_E^2\,Q^2 + \frac{b_2}{120}R_E^4\,Q^4
 - \frac{b_3}{5040}R_E^6\,Q^6.$$
The coefficients are given by models where form factor and charge
distributen can be expressed in terms of elementary functions with one
parameter $R_E$ and the expected values $<r^n>$ are simple multiples
of $R_E^n$. We have put the relevant formulas in the appendix
\ref{sec:appendix}. However, the authors of \cite{Griffioen:2015hta}
limited their analysis to three models, i.e., exponential, Gaussian
and box shaped charge distribution and they did not investigate the
bias (Section~\ref{sec:bias}) and the robustness
(Section~\ref{sec:robust}) of their ansatz. We will show that this
is a severe shortcoming that completely invalidates their conclusion.

The exponential or dipole model is of course an obvious choice for the
proton. The other two form factor models have a smaller kurtosis than
the dipole and would be suitable for light and heavy nuclei,
respectively. Therefore we look at two more models: Yukawa I and
II. Both are more ``peaked'' than the dipole model and the later, a
simple pole, has been used to fit the pion form factor.

In order to evaluate the bias and the robustness of the five models we
generated pseudo data equally spaced in the momentum transfer range
$Q^2=\unit[(0.004\ldots 0.02)]{(GeV/c)^2}$ with a constant standard
deviation of $0.5\%$, 201 data points in total. The result of this
analysis is shown in Tab.\ \ref{tbl:curvature}. With a few exceptions,
any mismatch between assumed functional form and actual functional
form leads to large biases. We conclude that, as long as one does not
regularly win the lottery, one should not guess the functional shape.

\begin{table}
\begin{tabular}{|l||r|r|r|r|r|} \hline
& \multicolumn{5}{|c|}{$b_{2/3}$ of fit function
  according to} \\ \hline
Input model& \multicolumn{1}{|c|}{Dipole}
& \multicolumn{1}{|c|}{Gauss}
& \multicolumn{1}{|c|}{Box}
& \multicolumn{1}{|c|}{Y.\ I}
& \multicolumn{1}{|c|}{Y.\ II} \\ \hline
\hline
Dipole & 0(4) & -5(4) & -9(4) & 22(5) & 5(4) \\ \hline 
Gauss & 5(4) & 0(4) & -3(4) & 28(5) & 11(4) \\ \hline 
Box & 9(4) & 3(4) & 0(4) & 31(5) & 14(4) \\ \hline 
Yukawa I & -21(4) & -26(4) & -29(4) & -1(5) & -16(5) \\ \hline 
Yukawa II & -5(4) & -11(4) & -14(4) & 16(5) & 0(5) \\ \hline 
P$\times$D & -8(4) & -14(4) & -17(4) & 13(5) & -3(4) \\ \hline 
Spline & -6(4) & -11(4) & -14(4) & 16(4) & -1(4) \\ \hline 
\end{tabular}
\label{tbl:curvature}
\caption{Bias and standard deviation in attometer. A mismatch between
  assumed functional form  and actual functional form can lead to
  significant biases. }
\end{table}

\subsection{$G_E/G_M$ ratio and continued fraction expansion}
In the second part of \cite{Griffioen:2015hta}, the authors extract
$G_E$ and $G_M$ from the whole Mainz data set using 
\begin{equation}
\label{gegmratio}
\mu_pG_E/G_M=1-Q^2/Q_0^2,
\end{equation}
with $Q_0^2=\unit[8]{(GeV/c)^2}$. The form of eq.\ \ref{gegmratio} is
motivated by measurements of the form factor ratio using
polarization. We believe that this approach is dangerous and wrong on
multiple accounts:
\begin{itemize}
\item It is a well known fact that the form factor ratio extracted
  from polarized measurements is different from the one extracted from
  unpolarized experiments. The most likely explanation is the neglect of  two-photon
  exchange, which affects mainly the unpolarized
  measurements. However, so far, this is only a conjecture. 
 The (unpolarized) Mainz dataset has no full two-photon exchange
 corrections applied. It is questionable to extract the form factors
 assuming a ratio from polarized data.
\item The linear fall-off describes the gross behaviour of the ratio
in  polarized data, but the world data set is certainly not good
enough to see structures beyond that, especially below $\unit[1]{(GeV/c)^2}$, where the current polarized data is somewhat in
disagreement with each other.
\end{itemize}

\begin{figure}
\includegraphics{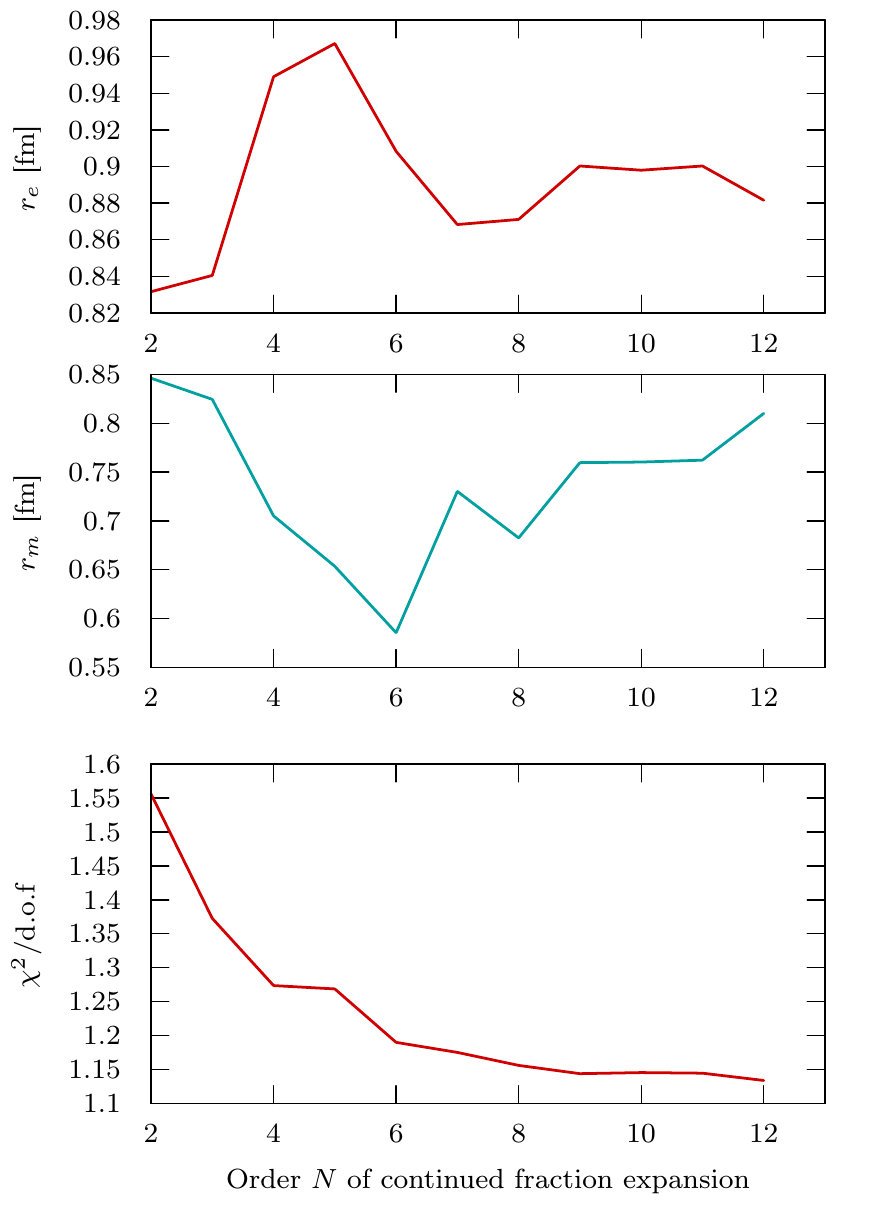}
\caption{\label{figcfe}Extracted radii and achieved
$\chi^2/\mathrm{d.o.f.}$ as a function of $N$ using continued fraction
expansions with $N$ parameters for $G_E$ and $G_M$. In contrast to
other models, the knee is very soft.  For fits with more than 3
parameters, $r_E>\unit[0.868]{fm}$.}
\end{figure}

The authors then fit their extracted $G_E$ using a continued fraction
expansion
\begin{equation}
G_E(Q^2)=\frac{p_1}{1+\frac{p_2Q^2}{1+{\frac{p_3Q^2}{1+...}}}}.
\end{equation}
with 4 parameters, they achieve a $\chi^2/\mathrm{d.o.f.}$ of 1.6 and
claim that the data are well-fit on average in all regions of $Q^2$. 
We can not follow this logic: a $\chi^2/\mathrm{d.o.f.}$ of 1.6 is
excessively high. 
Using our standard approach used in the Mainz analysis, we fit a
continued fraction expansion of both $G_E$ and $G_M$. The fits proved
difficult, with many local minima, and we can not rule out that better
solutions exist. Nevertheless, the results, shown in
Fig.\ \ref{figcfe}, are interesting. At order 4, we already achieve a
red.\ $\chi^2$ substantially better than 1.6. Order 5 is only
marginally better---we suspect a better solution exists, but our fit
fails to find it, despite randomizing the starting values (see Section \ref{sec:minimum}). At higher orders, we again see a substantial
gain. Around 9, the red.\ $\chi^2$ is comparable to the best models of
our earlier analysis, with a somewhat larger $r_e\approx \unit[0.899]{fm}$.
For fits with more than 3 parameters, our extracted radius is
always larger than $\unit[0.868]{fm}$. 
It is unclear whether the difference to \cite{Griffioen:2015hta} is
explained alone by the different extraction method. It is possible
that their fitting algorithm falls victim to the adverse conditions of
the fit too. Comparing their result for a double dipole (red.\
$\chi^2=1.6$) and our (1.29), the former seems likely.
N.B.: we can not quite follow their remark about smoothly and
monotonically falling fit functions. All our fit functions are smooth
and monotonic, and achieve $\chi^2$ around 1.15. 

\subsection{Dipole fit to low-Q data}
\begin{figure}
\includegraphics{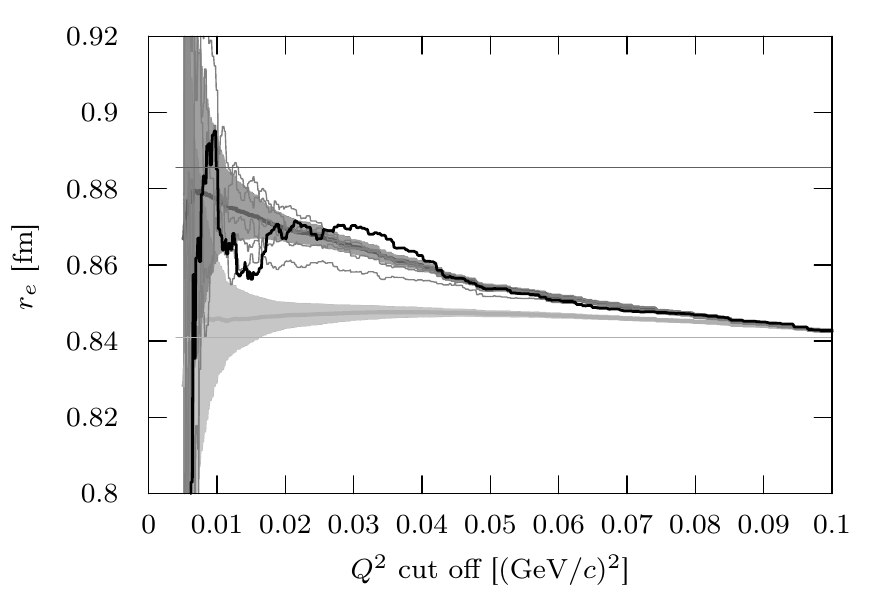}
\caption{\label{dipfit}Dipole fit to (pseudo)-data, same nomenclature
  as Fig.\ \ref{fofit}. At a  cut-off of $\unit[0.1]{(GeV/c)^2}$, a
  dipole fit extracts almost identical values when fit to the two
  pseudo-data samples, in agreement with the value extracted from
  data. However, for lower cut-off values, the radii extracted from
  data replicate the behavior of the pseudo-data sample with
  $r_e=\unit[0.8855]{fm}$, and does not follow the one with a small radius.}
\end{figure}

In \cite{Horbatsch:2015qda}, Horbatsch and Hessels compare a conformal mapping
polynomial fit with a dipole fit, for a range of $Q^2$ cut-offs and
orders. Their z-expansion fits exhibit indications of the problems
described in Sections \ref{sec:conformal}, but generally reproduce the large radius, in
agreement with our findings and in stark contrast to
\cite{Lorenz:2014vha}. Their dipole fit, for data up to $\unit[0.1]{(GeV/c)^2}$,
 yields a value of $\unit[0.842(2)]{fm}$. While this might
puzzle the reader, this is completely expected: the dipole model is
known to have a strong bias, as already demonstrated in \cite{Bernauer:2013tpr} for the whole data set. 
We repeat the procedure described in Section \ref{pseudodata}, fitting a
dipole model. The results are shown in Fig.\ \ref{dipfit}.  At $\unit[0.1]{(GeV/c)^2}$, the extracted radii are identical, and no decision
can be made. At lower cut-offs, the data clearly prefer the
pseudo-data sample with a large radius. It is worthwhile to note that
the dipole fit to the pseudo data sample with the large radius has a
negative bias larger than the expected statistical error for cut-offs
larger than $\unit[0.01]{(GeV/c)^2}$. A reliable extraction of the
radius can therefore not be expected.

\subsection{Statistical methods to decide order}
\label{sec:occamsbeard}
In \cite{Higinbotham:2015rja}, the authors use the F-test to decide which order
of polynomials to use. Besides the points addressed in Section \ref{sec:tests}, the statistical interpretation is flawed on a very basic level: they reference a critical value of $4.3$ for CL=$95\%$, which is the critical value for the rejection of H0 at this level, i.e., with an F-test value higher than $4.3$, one should reject the simpler model, with a $5\%$ probability that the rejection is wrong. They however claim that their value below this threshold rejects H1, i.e., the more complex model, at $95\%$ CL. This inversion can of course not be done, and indeed, no confidence level can be given for this type of error easily, because the nominator in the F-test does not follow a standard Fisher-Snedecor distribution anymore, and because one is not restricted to just one higher order.

Nevertheless, with the pseudo data groups above, we can
easily test what their flawed method would produce:

Comparing first and second order fits, the F-test would prefer (i.e., not rule out at CL=95\%) the linear model up to
$0.015$ (large $r_e$ group) and $\unit[0.02]{(GeV/c)^2}$ (small $r_e$), respectively. At
these $Q^2$,  the bias of the linear model is $\unit[0.03]{fm}$ and
$\unit[0.02]{fm}$, as can be seen in Fig.\ \ref{fofit}. 

\begin{figure}
\includegraphics{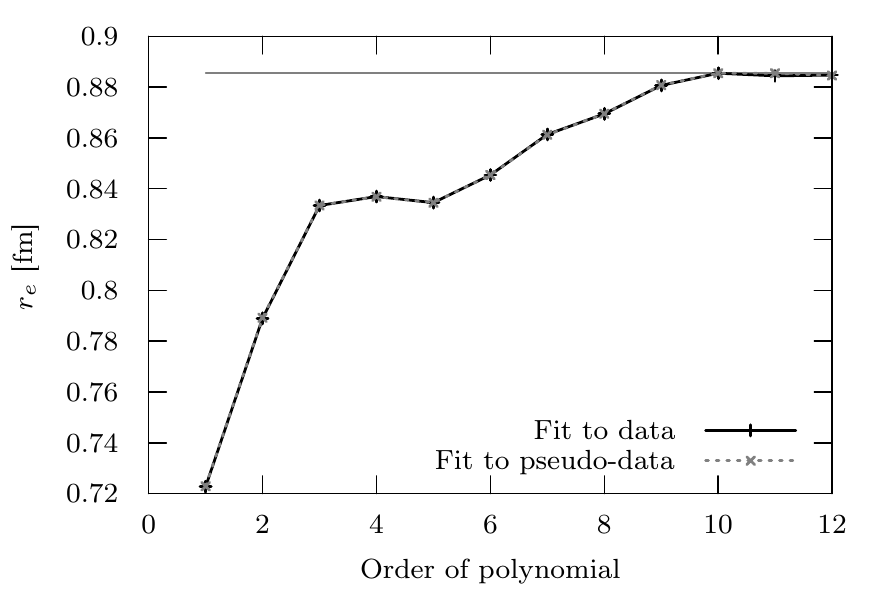}
\caption{\label{ptolemaeic} Dependency of the extracted radius from polynomial models on the order, for the full Mainz data set. Fits to pseudo data track the behavior of the fits to real data, and can only recover the input radius (indicated by horizontal line) at high orders. Both F-test and AIC reject models with orders $<9$.}
\end{figure}

For second and third order fits, the cross-over is around $\unit[0.045]{(GeV/c)^2}$ for the large radius pseudo data group, and above
$0.1$ for the smaller radius group (outside our simulation range),
albeit with a certain fraction of the individual data sets hitting the
threshold around $0.09$.  Comparing to Fig.\ \ref{sofit}, the method seems to work in this case
for the smaller radius group---the second and third order coefficients
of the input model are significantly smaller than for the large radius input model. For the larger radius
pseudo data group, however, the fit to pseudo data has a bias of
$\unit[0.012]{fm}$ and $\unit[0.045]{fm}$, and the fit to data up this
point produces a {\it large} radius.

Let us now look at the behavior of fits of different order to the full data set. For this test, we will again use our \nth{10} order polynomial as basis for the generation of pseudo data. We then fit polynomial models with different orders both to the real data and to the pseudo data. Instead of relying on a fixed $G_M$ fit as we did for the low-Q fits, we fit both $G_E$ and $G_M$ at the same time, repeating our approach of \cite{Bernauer:2010wm,Bernauer:2013tpr}. The results are shown in Fig.\ \ref{ptolemaeic}. The fits to the pseudo data replicate almost exactly the behavior of the fits to data, with also similar behavior for the F-test. It is interesting to see that lower orders linger around the muonic radius. However, this is a good example how to use statistical tests properly: The F-test rejects order 9 and below in favour of order 10, with a false rejection probability of $<5\%$. The Akaike information criterion accepts order 9, and has a minimum at order 10. It follows that the H0 hypothesis, the lower order models are correct, is rejected by the data.

We further want to note that Table III of \cite{Higinbotham:2015rja} is not consistent with its description. The listed values for $\chi^2$ and $\chi^2/\nu$ indicate that $\nu=N-j$, instead of $\nu=N-j-1$ given in the description---and even assuming that, there seems to be a rounding error.

\section*{Conclusion}
In summary, we inspected several recent refits of the Mainz data set
which result in small radii and found flaws of various kinds in all of
them. While a reanalysis of the data can not rule out faulty data---which would invalidate any extraction---we believe that the solution
of the puzzle can not be found in the fit procedure. We urge anybody
in the business to test their method using pseudo data generated from
the Mainz fits.

\section*{Acknowledgement}
We thank J\"org Friedrich, Kees de Jager and Thomas Walcher for helpful discussions.

\bibliography{note}

\appendix
\section{Form factors and charge distribution of selected models}
\label{sec:appendix}
In the Breit frame electric (and magnetic) form factors can be associated with the charge (and magnetic current) density distributions through a Fourier transformation:
\begin{align}
G(q) =& 4\pi \int^\infty_0 r^2 \, \rho(r) \,
         \sin\left(\frac{q \, r}{\hbar c}\right) \frac{\hbar c}{q \, r} \,\text{d}r \\
\rho(r)  =& \frac{4\pi}{(2\pi\,\hbar c)^{3}} \int^\infty_0 q^2 \, G(q) \,
             \sin\left(\frac{q \, r}{\hbar c}\right) \frac{\hbar c}{q \, r} \,\text{d}q
\end{align}
This implies that the electric form factor can be expanded in terms of $Q^2$ where the coefficients are the multiples of the expected values of $r^{2n}$ of the charge distribution:
\begin{align}
G(Q^2) =& \sum_{n=0}^\infty \frac{(-1)^n}{(2n+1)!} \,
           \langle r^{2n} \rangle\,Q^{2n} \\ \nonumber
=& 1 - \frac{\langle r^2 \rangle}{6}\,Q^2
    + \frac{\langle r^4 \rangle}{120}\,Q^4
    - \frac{\langle r^6 \rangle}{5040}\,Q^6 + \ldots
\end{align}

The Zemach moments of the nuclear charge distributions (see
\cite{Distler:2010zq} and references therein) are defined by

\begin{equation}
\langle r^n \rangle_{(2)} = \int d^3r \,r^n \rho_{(2)}(r)
\label{eq:zemach1}
\end{equation}

where $\rho_{(2)}(r)$ is the convolution of the charge distribution
\begin{equation}
\rho_{(2)}(r)= \int d^3r_2 \, \rho(|\vec{r}-\vec{r_2}|) \, \rho(r_2).
\label{eq:zemach2}
\end{equation}

The first and the third Zemach moment can also be expressed in
momentum space:

\begin{align}
\langle r \rangle_{(2)}  =& 
-\frac{4}{\pi} \, \int_0^{\infty} \frac{dQ}{Q^2}\left(G_E^2(Q^2)-1\right)
\\[2ex]
\langle r^3 \rangle_{(2)}  =& 
\frac{48}{\pi} \, \int_0^{\infty} \frac{dQ}{Q^4}
\left(G_E^2(Q^2)-1+\frac{Q^2}{3}\langle r^2 \rangle\right) \nonumber.
\label{eq:zemach3}
\end{align}

In the following sections we will give the form factors, the density
distributions and their expected values of $r^4$ and $r^6$ for
selected models as a function of $R=\sqrt{\langle r^2 \rangle}$. The
first and the third Zemach moment and Zemach's convoluted density are
shown as well. The latter is not available in closed form for the
Yukawa I model.
\allowdisplaybreaks

\subsection{Exponential (dipole) model}
\begin{align}
G(q)=& \left(1+\frac{1}{12}
        \left(\frac{q R}{\hbar c}\right)^2\right)^{-2} \nonumber \\
\rho(r)=&\frac{3 \sqrt{3}}{\pi R^3}
           \exp\left[-2\sqrt{3}\frac{r}{R} \right] \nonumber \\
\rho_{(2)}(r)=&\frac{3 \sqrt{3}}{8\pi R^5}
                 \left( 4r^2+2\sqrt{3}r R+R^2 \right)
 \nonumber \\ & \times
           \exp\left[-2\sqrt{3}\frac{r}{R} \right] \nonumber \\
\langle r^4 \rangle=&\frac{5}{2}\,R^4 \nonumber \\
\langle r^6 \rangle=&\frac{35}{3}\,R^6 \nonumber \\
\langle r \rangle_{(2)} =& \frac{35}{16\sqrt{3}}\,R \nonumber \\
\langle r^3 \rangle_{(2)} =& \frac{35\sqrt{3}}{16}\,R^3
\end{align}

\subsection{Gaussian}  
\begin{align}
G(q)=&\exp\left[-\frac16\left(\frac{q R}{\hbar c}\right)^2\right] \nonumber \\
\rho(r)=& \left( \sqrt{\frac{3}{2\pi}} \frac{1}{R} \right)^3
           \exp \left[ -\frac32\frac{r^2}{R^2} \right] \nonumber \\
\rho_{(2)}(r)=& \left( \sqrt{\frac{3}{\pi}} \frac{1}{2 R} \right)^3
                 \exp \left[ -\frac34\frac{r^2}{R^2} \right] \nonumber \\
\langle r^4 \rangle=&\frac{5}{3}\,R^4 \nonumber \\
\langle r^6 \rangle=&\frac{35}{9}\,R^6 \nonumber \\
\langle r \rangle_{(2)} =& \frac{4}{\sqrt{3\pi}}\,R \nonumber \\
\langle r^3 \rangle_{(2)} =& \frac{32}{3\sqrt{3\pi}}\,R^3
\end{align}


\subsection{Uniform}
\begin{align}
G(q) =& \left( \frac35\frac{\hbar c}{q R} \right)^2
         \left( - 5 \cos\left[ \sqrt\frac53 \frac{q R}{\hbar c}
         \right] \right. \nonumber \\
& \left. + \sqrt{15} \frac{\hbar c}{q R} \sin\left[ \sqrt\frac53
   \frac{q R}{\hbar c} \right] \right) \nonumber \\
\rho(r) =& \frac{3}{4\pi R^3} \left( \frac35 \right)^{3/2}
            \Theta\left[ \sqrt\frac35 R - r \right] \nonumber \\
\rho_{(2)}(r) =&  \frac{27}{8000\pi R^6} \,
     \Theta\left[ 2\sqrt\frac35 R - r \right] \nonumber \\
  & \times \left( 3 r^3 - 60 r\, R^2 + 80 \sqrt\frac53\, R^3 \right)
 \nonumber \\
\langle r^4 \rangle =& \frac{25}{21}\,R^4 \nonumber \\
\langle r^6 \rangle =& \frac{125}{81}\,R^6 \nonumber \\
\langle r \rangle_{(2)} =& \frac{12}{7}\sqrt\frac35\,R \nonumber \\
\langle r^3 \rangle_{(2)} =& \frac{160}{63}\sqrt\frac53\,R^3
\end{align}

\subsection{Yukawa I}
\begin{align}
G(q)=& \sqrt{2} \frac{\hbar c}{q R}
        \arctan\left( \sqrt\frac12 \frac{q R}{\hbar c} \right) \nonumber \\
\rho(r)=& \frac{1}{2 \sqrt{2} \pi r^2 R}
           \exp\left[ -\sqrt{2} \frac{r}{R} \right] \nonumber \\
\langle r^4 \rangle=&6\,R^4 \nonumber \\
\langle r^6 \rangle=&90\,R^6 \nonumber \\
\langle r \rangle_{(2)} =& \frac13\sqrt2(1+2\log[2])\,R \nonumber \\
\langle r^3 \rangle_{(2)} =& \frac35\sqrt2(3+4\log[2])\,R^3
\end{align}

\subsection{Yukawa II}
\begin{align}
  G(q) =& \left( 1 + \frac16 \left( \frac{q R}{\hbar c} \right)^2\right)^{-1}\nonumber\\
  \,&\,\nonumber\\[-1ex]
\rho(r) =& \frac{3}{2 \pi r R^2}\exp\left[ -\sqrt{6} \frac{r}{R} \right] \nonumber \\
\rho_{(2)}(r) =& \frac{3}{2 \pi}\sqrt\frac32\frac{1}{R^3}
                  \exp\left[ -\sqrt{6} \frac{r}{R} \right] \nonumber \\
\langle r^4 \rangle=&\frac{10}{3}\,R^4 \nonumber \\
\langle r^6 \rangle=&\frac{70}{3}\,R^6 \nonumber \\
\langle r \rangle_{(2)} =& \sqrt\frac32\,R \nonumber \\
\langle r^3 \rangle_{(2)} =& 5\sqrt\frac23\,R^3
\end{align}

\end{document}